\documentclass[preprint,iop]{emulateapj}
\usepackage{apjfonts}
\usepackage{amsmath}
\usepackage{natbib}
\usepackage[colorlinks=true,linkcolor=blue,citecolor=blue]{hyperref}

\begin{document}

\newcommand{\beq}{\begin{equation}}
\newcommand{\eeq}{\end{equation}}
\newcommand{\beqar}{\begin{eqnarray}}
\newcommand{\eeqar}{\end{eqnarray}}
\newcommand{\e}{\varepsilon}
\newcommand{\rt}{r_{\rm t}}
\newcommand{\rs}{r_{\rm s}}
\newcommand{\Mbh}{M_{\rm bh}}
\newcommand{\rp}{r_{\rm p}}
\newcommand{\risco}{r_{\rm isco}}
\newcommand{\rb}{r_{\rm b}}
\newcommand{\tdyn}{t_{\rm dyn}}
\newcommand{\tfb}{\tau_{\rm fb}}
\newcommand{\tpeak}{\tau_{\rm peak}}
\newcommand{\Jc}{J_{\rm c}}
\newcommand{\Jlc}{J_{\rm lc}}
\newcommand{\tr}{t_{\rm r}}
\newcommand{\tJ}{\tau_{\rm J}}
\newcommand{\Flc}{F_{\rm lc}}
\newcommand{\dmA}{\Delta m_{\rm A}}
\newcommand{\dmB}{\Delta m_{\rm B}}
\newcommand{\Ledd}{L_{\rm Edd}}
\newcommand{\torb}{\tau_{\rm orb}}
\newcommand{\tevol}{\tau_{\rm evol}}
\newcommand{\tcoll}{\tau_{\rm coll}}
\newcommand{\te}{\tau_{\e}}

\title{Spoon-Feeding Giant Stars to Supermassive Black Holes: Episodic Mass Transfer From Evolving Stars and Their contribution to the quiescent activity of  galactic nuclei} 
\author{Morgan MacLeod, Enrico Ramirez-Ruiz, Sean Grady, and James Guillochon}
\shortauthors{}
\affil{Department of Astronomy and
  Astrophysics, University of California, Santa Cruz, CA
  95064}
\email{mmacleod@ucolick.org}

   
\begin{abstract} 
Stars may be tidally disrupted if, in a single orbit, they are scattered too close to a supermassive black hole (SMBH). Tidal disruption events are thought to power luminous but short-lived accretion episodes that can light up otherwise quiescent SMBHs in transient flares. Here we explore a more gradual process of tidal stripping where stars approach the tidal disruption radius by stellar evolution while in an eccentric orbit. After the onset of mass transfer, these stars episodically  transfer mass to the SMBH every pericenter passage giving rise to low-level flares that repeat on the orbital timescale. Giant stars, in particular, will exhibit a runaway response to mass loss and ``spoon-feed'' material to the black hole for tens to hundreds of orbital periods. In contrast to full tidal disruption events, the duty cycle of this feeding mode is of order unity for black holes $\Mbh \gtrsim 10^7 M_\odot$. This mode of quasi-steady SMBH feeding is competitive with indirect SMBH feeding through stellar winds, and spoon-fed giant stars may play a role in determining the quiescent luminosity of local SMBHs. 
\end{abstract}

\maketitle

\section{Introduction}

The prevalence of quasar activity at early epochs provides evidence that supermassive black holes (SMBHs) must lurk in the centers of many galactic halos \citep{Soltan:1982vm}. Yet, in the local universe the vast majority of galactic center SMBHs exhibit little activity.  Recent study has revealed that many of these SMBHs are likely shining due to mass accretion, but only at a tiny fraction of their Eddington luminosities, $L/\Ledd \ll 1$  \citep{Ho:2009fq}.  To best understand the origin of the low observed Eddington ratios of local SMBHs, it is important to develop a census of the processes that combine to establish a minimum, ``floor'', feeding level,  $\dot M$. This floor accretion rate determines the most typical level of SMBH activity and therefore gives rise to the quiescent luminosity of galactic nuclei, where $L=\eta \dot M c^2$. In galactic nuclei devoid of gas, any potential fuel comes solely from the dense stellar clusters that surround SMBHs, thus stars alone serve to establish a lower limit of SMBH activity. By constructing an accurate census of fuel sources arising from the stellar distribution, we can eventually constrain the accretion efficiency $\eta$ in an effort to better understand the accretion flows onto these SMBHs. 

Stars feed the black hole in two primary ways, directly and indirectly. Indirect feeding arises from processes that inject material into the nuclear cluster medium, as occurs with stellar winds and stellar collisions \citep{Holzer:1970eo,Coker:1997cg,Loeb:2004bs,Quataert:2004bp,Cuadra:2005hu,Cuadra:2006ju,Cuadra:2008bd,Freitag:2002gw,Volonteri:2011hj,Rubin:2011jt}. To reach the SMBH, material fed indirectly into the cluster medium must overcome a further barrier to accretion in the form of feedback from the stars themselves and the SMBH \citep[e.g.][]{Blandford:1999dp,Quataert:2004bp,Shcherbakov:2013ip}. 

Direct feeding of the SMBH results from tidal interactions between stars and the SMBH. Tidal interactions result in a dynamically assembled disk \citep[e.g.][]{Bogdanovic:2004gd,Guillochon:2013vh}, which is relatively invulnerable to the feedback processes which plague our understanding of indirectly fed accretion mechanisms.  Stars passing within approximately a tidal radius of the SMBH, $\rt \equiv (\Mbh/M_*)^{1/3} R_*$, where $M_*$ and $R_*$ are the stellar mass and radius, will experience strong tidal distortions and may be partially or completely destroyed by the black hole's tidal field \citep[e.g.][]{Hills:1975kh,Rees1988}. Half of the tidally stripped debris of tidal disruption eventually falls back to the SMBH, forms a disk, and viscously accretes. Full tidal disruptions of main-sequence \citep{Guillochon:2013jj} and giant stars \citep{MacLeod:2012cd} produce luminous flares \citep[e.g.][]{Evans:1989ju,Strubbe:2009ek,Strubbe:2011iw,RamirezRuiz:2009gw,Lodato:2009ib,Guillochon:2013vh}, but the duration of flares is generally short compared to their repetition time, $\sim 10^{4}$ yr \citep{Rees1988,Magorrian:1999fd,Wang:2004jy}.  In quiescence, the accretion rate to the SMBH is determined by the late time fallback of tidal debris \citep{Milosavljevic:2006jj}, which decays roughly as $\dot M \propto t^{-5/3}$. While the average accretion rate is relatively large, $\sim M_\sun / t_{\rm repeat} \sim 10^{-4} M_\sun \text{yr}^{-1}$, the rapid decline in the fallback after peak results in a median accretion rate that is much lower, $\sim \dot M_{\rm peak} (t_{\rm repeat}/t_{\rm peak})^{-5/3} \sim 10^{-9} M_\sun \text{yr}^{-1}$, assuming typical parameters for a main-sequence star. 

In this paper, we study a mechanism that does not result in luminous flares but can fill in between the tails of tidal disruption events and result in much higher median accretion rates. This process is the mass transfer that ensues when a giant star grows, over the course of many orbital periods, such that its tidal disruption radius becomes comparable to its orbital pericenter distance.  Because of the large disparity between $\rt$ for a main-sequence star and $\rt$ for a giant star, there exist many main-sequence star orbits that pass safely within the giant star tidal radius at pericenter. While on the main sequence, a star in such an orbit experiences little disturbance from the black hole's tidal field. However, as the star evolves off of the main sequence it expands, and, as a result, its mean density drops. With each passing orbit, the star therefore feels the tidal forcing from the SMBH with increasing strength. Eventually, the star is distorted to the point that  a fraction of its envelope mass is removed at pericenter.  

As the star evolves up the giant branch, its recently developed dense core helps protect it against complete disruption \citep{Hjellming:1987ci,MacLeod:2012cd,Liu:2012er}, and the surviving remnant  therefore returns to pericenter after each orbital period. The adjustment of the star's structure to the mass loss it undergoes determines the strength of these subsequent encounters and the number of orbits over which the giant's envelope is depleted.  
Stars that undergo many passages by the SMBH are altered by these encounters \citep{Alexander:2003dp,Alexander:2003jr,Alexander:2005ij,Li:2013ij}, and the star's history of encounters with the SMBH will determine the nature of the subsequent passages. 
It is worth noting that an orbital history where the star returns to pericenter many times is distinct from the single-passage encounters that have received the majority of focus in previous studies of tidal disruption in galactic nuclei. Recent studies of the tidal disruption of objects on eccentric orbits have looked at giant planets \citep{Guillochon2011}, repeating flares from stars deposited into tightly bound orbits through binary disruption \citep{Antonini:2011ia}, and the fallback properties of tidal debris in eccentric disruptions \citep{Hayasaki:2012ec}. 

Giant stars that repeatedly transfer small amounts of their envelope mass to the SMBH (which we call ``spoon-feeding'') do so over many orbital periods. As a result, this channel of SMBH feeding results in a quasi-steady feeding rate to the black hole, in contrast to the highly peaked feeding due to the tidal disruption of stars. We find that as a result of effectively spreading the bulk of their mass over longer feeding timescales than typical tidal disruption events, spoon-fed giant stars may play a significant role in determining the quiescent luminosity of local SMBHs. The feeding that results from these mass-transferring stars is competitive with the amount of mass fed indirectly to the SMBH by stellar winds. 

This paper is organized as follows. 
In Section \ref{sec2}, we discuss the onset of mass transfer resulting from the the evolution of a giant star trapped in an elliptical orbit and the SMBH. 
In Section \ref{sec3}, we show that the star episodically spoon-feeds mass to the SMBH over the course of many pericenter passages. 
 In Section \ref{sec4}, we estimate the expected population of these trapped stars and estimate the rate at which they evolve to  feed mass to the SMBH.
 In Section \ref{sec5}, we discuss the effects of these mass-transfer events on the floor activity level and duty cycle of local, tidally-fed SMBHs. 
In Section \ref{sec6}, we conclude and offer prospects for future study.

\section{Mass transfer from evolving stars}\label{sec2}
A main-sequence star in an orbit that passes within the maximum red giant tidal radius at pericenter may survive for a long time relatively unperturbed by the black hole.  
Eventually, the star leaves the main sequence and evolves up the giant branch, at which point its radius expands and its mean density drops. At each pericenter passage the evolving star feels the tidal force of the black hole with increasing strength. Finally, the star grazingly begins to lose mass at pericenter. In this section we present a hydrodynamical simulation of this first disruptive passage. We will use this simulation to study the effects of the encounter on the surviving stellar core and to motivate a semi-analytical model for the subsequent passages in Section \ref{sec3}.

Previous analytic work and numerical simulations of mass transfer episodes in eccentric binaries have focused primarily on the context of stellar mass binaries \citep{Regos:2005gi,Sepinsky:2007fp,Sepinsky:2009fg,Lee:2010in,Sepinsky:2010ga,Lajoie:2010jv,East:2012hg,East:2012es,Davis:2013vj}. Recently, \citet{faber2005} and \citet{Guillochon2011} have numerically explored higher mass ratio eccentric encounters in the context of the orbital dynamics and disruption of giant planets in eccentric orbits about their parent stars. 
\citet{Antonini:2011ia} have speculated about the fate of stars that are dynamically deposited on tightly bound orbits through binary star disruptions, while \citet{Hayasaki:2012ec} and \citet{Dai:2013un} have numerically studied the fallback properties of tidal debris in eccentric disruptions.

\begin{figure}[tbp]
\centering
\includegraphics[width=\columnwidth]{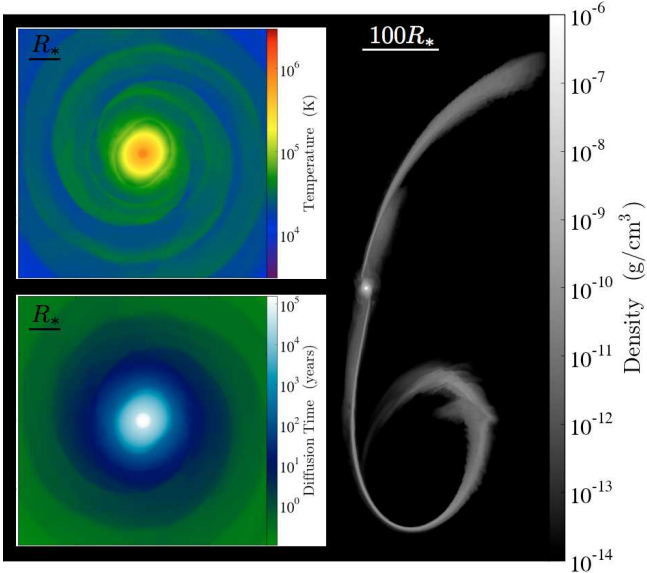}
\caption{Mass stripping from a star in an eccentric orbit around a SMBH. The main panel shows the formation of black hole bound and unbound tidal tails. Bound material streams back pericenter where it will circularize and drain into the SMBH. The upper inset shows a layer of stellar envelope heated by spiral shocks that originate from the remnant's interaction with material falling back from the tidal tails. The lower inset shows that despite this heating, the photon diffusion time through these tenuous layers is short enough that the envelope cools in much less than a typical orbital period, $\torb$. Our initial stellar model is a nested polytrope representative of a $1.4 M_\odot$, $50 R_\odot$ red giant with a $0.3 M_\odot$ dense core. The simulation shown was computed at a smaller mass ratio $(\Mbh = 10^4 M_\odot)$ and with lower eccentricity $e=0.8$ than the encounters described in the text in order to illustrate the fallback and circularization processes. The scalebar is in units of $R_* = 50R_\odot$ in this case.  }
\label{fig1}
\end{figure}

In Figure \ref{fig1}, we present a simulation of a grazing encounter between a giant star and the SMBH preformed in the {\tt FLASH} hydrodynamics code \citep{Fryxell:2000em}  using the method described in detail in \citet{MacLeod:2012cd}. Our formalism is based on the {\tt FLASH4} code in Newtonian gravity and follows  the encounter in the frame of the star \citep{Guillochon:2009di,Guillochon2011,MacLeod:2012cd,Liu:2012er,Guillochon:2013jj}. Our initial stellar model is a nested polytrope representative of a $1.4 M_\odot$, $50 R_\odot$ red giant with a $0.3 M_\odot$ dense core. A core mean molecular weight of twice that of the envelope fluid produces a relatively inert core. The structure of both the core and envelope are $n=1.5$ polytropes. The adiabatic fluid gamma is $\Gamma = 5/3$ for the envelope gas, and it is $\Gamma = 5$ to model the tidally unperturbed core. The star is initially resolved by 90 grid cells in radius. After the encounter, grid refinement adaptively follows the density of the stripped gas. 

Even in a relatively grazing encounter, the star is subject to a rapidly time-varying potential at pericenter \citep{Regos:2005gi}. It is non-linearly distorted and some portion of its envelope mass may be unbound from the stellar core. This material is ejected from the star in two tidal tails, one of which is bound to the black hole, while the other is ejected on hyperbolic trajectories. The amount of mass lost depends on the impact parameter of the encounter, which is defined by the ratio of  the star's tidal radius to the pericenter of its  orbit, $\beta \equiv \rt / \rp =  ( R_*/ \rp) (\Mbh/M_*)^{1/3}$. The giant star in Figure \ref{fig1} encounters the black hole with $\beta = 0.6$ and loses $\Delta M \approx 10^{-2} M_\sun$.   A linearized approach to determining the degree of mass loss at pericenter by calculating the degree to which the stellar envelope overflows its effective Hill sphere at pericenter, $\rp (M_*/\Mbh)^{1/3}$, would suggest that no mass is lost at these grazing $\beta$, where the star is still a factor of $\sim 2$ smaller than its  Hill sphere. Thus it is extremely important to account for the non-linear distortion of the star, even when the degree of mass loss is very small. We discuss a simple model for the degree of mass loss as a function of $\beta$ in Section \ref{sec3}.

Some of the material originally ejected into the tidal tails during the encounter will eventually fall back to the stellar surface. The insets of Figure \ref{fig1} show the state of the remnant post-encounter. As a fraction of the stripped material falls back to the oscillating and rotating stellar envelope, spiral shocks are generated. These shocks heat a tenuous layer of envelope material with mass similar to the fallback mass ($\lesssim \Delta M$) that adiabatically expands to extend significantly beyond the initial stellar radius (upper inset panel).  By contrast, the interior portion of the star, $r\ll R_*$, is not heated by shocks in encounters where $\Delta M \ll M_*$. Our adiabatic simulation does not capture the radiative cooling of this material which extends to beyond the initial stellar radius. The lower inset panel of Figure \ref{fig1} shows that the local photon diffusion time (approximated as $\tau_{\rm diff} \sim \rho \kappa_{\rm es} R_*^2/c$) through these outermost heated layers is very short, much less than an orbital period, which we will denote $\torb$. We therefore expect these outermost layers to cool effectively despite the heating due to interaction with the fallback from the debris streams.  As a result, this fallback heating should be a small perturbation to the remnant's structure. 

Additional heat may be deposited into the stellar interior through the dissipation of oscillation energy or interaction with gas in the circum-black hole medium. However, the tidal heating effect has been studied in detail by \citet{Li:2013ij} and was shown to be a small perturbation to stellar structure for timescales comparable to the star's red giant branch lifetime. This is partially because heat deposited into a giant star's envelope (rather than its core) is easily radiated due to the short diffusion time through the envelope \citep{1987ApJ...318..261M}. Interactions between the star and the remnant disk at pericenter may also heat the stellar envelope through shocks \citep[e.g.][]{Armitage1996,Dai:2010bf}. In the case considered here, this effect is likely to be of small importance because the disk mass will typically only be of order $10^{-2} M_\odot$ (See Figure \ref{fig2}), spread to extremely low density over the tidal sphere. These factors suggest that the state of the star in its subsequent encounters with the SMBH will be dominated by the stellar structure's  response to the mass loss  experienced, rather than the effects of extra heating or orbital evolution. 

Following a passage by the SMBH, changes in the remnant's orbit will alter the properties of subsequent encounters. Of particular importance in determining the strength of the encounter is the orbital angular momentum, which determines the pericenter distance. There are several effects which can potentially modify the orbital energy and angular momentum of the remnant. First, due to the cumulative effect of encounters with other stars in the stellar cusp around the SMBH, the remnant's orbit undergoes a random walk in orbital energy and angular momentum. In Section \ref{sec4}, we define the phase space of stellar orbits for which this random walk is small. 

Second, any asymmetry in the mass ejection between the tidal tails results in a change in the orbital energy of the surviving remnant \citep{faber2005,Guillochon2011,Liu:2012er,Cheng:2013cm,Manukian:2013ce}. This change in energy maximizes around the star's own specific binding energy, $E_* \approx G M_*/R_*$, for coreless stars and deep encounters \citep{Cheng:2013cm,Manukian:2013ce}, but it is strongly limited by the presence of a stellar core, as is the case in giant planets \citep{Liu:2012er}.

Finally, non-radial oscillations are excited in the remnant following the encounter leading to a transfer of orbital energy and angular momentum into stellar oscillation energy and angular momentum. The magnitude of these perturbations are typically a fraction of the star's binding energy or breakup angular momentum. This  result was analytically predicted by \citet{1977ApJ...213..183P}, and has more recently been numerically explored in the case of objects without \citep{Guillochon2011,Cheng:2013cm} and with \citep{Liu:2012er} cores.  In the case of giant stars interacting with SMBHs on bound orbits, the star's orbital energy and angular momentum are both large compared to the giant star's binding energy and maximum rotational angular momentum. Typical values for these ratios of orbital binding energy to stellar binding energy are
\beq
\frac{E_{\rm orb}}{E_{*}} \approx \frac{R_*}{a} \frac{\Mbh}{M_*} \approx 11 \left(\frac{R_*}{ 50 R_\sun} \right) \left(\frac{a}{1 \text{pc}} \right)^{-1} \left(\frac{\Mbh}{10^7 M_*} \right),
\eeq
where $a$ is the orbital semi-major axis, and $E_{\rm orb} = G \Mbh / (2 a)$. By a similar analysis, the ratio of the orbital angular momentum, $J_{\rm orb} \approx \sqrt{2 G \Mbh \rp}$, to breakup rotational angular momentum of the star, $J_* \approx \sqrt{G M_* R_*}$, is of order 
\beq
\frac{J_{\rm orb}}{J_{*}} \approx \left(\frac{\Mbh}{M_*} \right)^{2/3} \approx 5\times 10^4 \left(\frac{\Mbh}{10^7 M_*} \right)^{2/3},
\eeq
if the substitution that the pericenter distance equals the tidal radius, $\rp=\rt$ is made (which leads to the lack of dependence on the stellar radius, $R_*$, in the above expression). Since the orbital quantities are much larger than the maximum reservoir of binding energy or rotational angular momentum available in the giant star, tidal excitation cannot induce substantial changes in the orbit.

Interestingly, all of these processes result in only small perturbations to the giant star's orbit. The orbital parameters of the bound giant star remnant are therefore essentially unchanged following the passage by the SMBH. As a result, in subsequent orbits the star will return to the same pericenter distance with a similar orbital period.

\section{ Episodic flares over many pericenter passages}\label{sec3}

In this section, we model the encounter history of a giant star on a bound orbit with the SMBH after the onset of mass transfer. 
Previously, we argued that the remnant's orbit is essentially unchanged by the encounter with the SMBH. 
Therefore, we can determine the mass lost each pericenter passage by calculating the changes in the stellar structure and comparing the pericenter of the star to its new tidal radius through the impact parameter, $\beta$. 
We adopt a model that combines an analytic description of the degree of mass loss and its return to the black hole with a stellar evolution calculation of the adjustment of the mass-losing star's structure. This approach is necessary to explore these multiple passage encounters because the range of timescales between the star's dynamical time and a typical orbital period make a full hydrodynamic calculation prohibitively computationally expensive. Recent work by \citet{Zalamea:2010eu} has similarly adopted an analytic model to study runaway flares from the progressive disruption of a white dwarf by an intermediate mass black hole.

\subsection{Mass Stripping}

To predict the degree of mass loss at each passage as a function of pericenter distance, we adopt a simple approximating formula motivated by simulation results from \citet{MacLeod:2012cd} and \citet{Guillochon:2013jj},
\beq\label{deltam}
\Delta M(\beta) = f(\beta) \left(   \frac{M_* - M_{\rm c}}{M_*}    \right)^2 M_* ,
\eeq
where $M_{\rm c}$ is the core mass and 
\beq
f(\beta) = \begin{cases}  
0 &  \mbox{if } \beta < 0.5 ,  \\
\beta/2 - 1/4 &  \mbox{if }  0.5 \leq \beta \leq 2.5 ,   \\
1 &  \mbox{if } \beta > 2.5    .
\end{cases}
\eeq
This parameterization captures two critical features of the simulation results. First, convective stars with condensed cores begin to lose mass around $\beta \sim 0.5$ -- a much more grazing encounter than linear models of the mass loss would predict. Yet, the tidal stripping does not reach its maximum until much deeper encounters, around $\beta \sim 2.5$. Second, while only the envelope material is susceptible to disruption, the increased influence of the core makes it more difficult to remove envelope material when the core is a larger mass-fraction of the star \citep[thus the squared dependence on $M_{\rm env}/M_{\rm c}$, see Figure 5 of][]{MacLeod:2012cd}.

Each mass loss episode results in a readjustment of the star's structure and therefore a new effective impact parameter with each pericenter passage. The importance of the adjustment of the mass-losing star's structure in the context of extreme mass ratio circular binaries has been demonstrated by \citet{Dai:2011tz} and \citet{Dai:2011tn}. We calculate the changes to the stellar properties using the {\tt MESA} stellar evolution code  \citep{Paxton:2011jf,Paxton:2013th}.\footnote{version 4849} Our stellar models are non-rotating, and the only source of mass loss is the interaction with the black hole. In the {\tt MESA} models, we allow the star to adjust to the mass loss continuously by applying an effective stellar wind that carries away the outermost envelope material at a rate $\dot M = \Delta M / \torb$, recalculated each pericenter. Timesteps are chosen such that each orbital period, $\torb$, is resolved by ten steps, but our results are not sensitive to this choice. 

\begin{figure}[tbp]
\begin{center}
\includegraphics[width=\columnwidth]{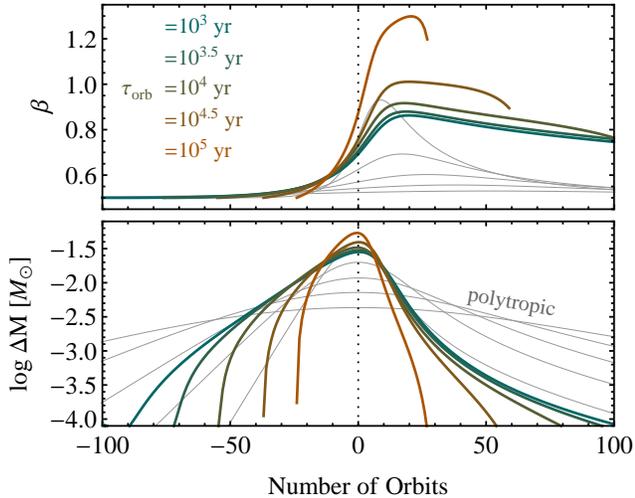}
\caption{Episodic mass transfer from a giant star to the SMBH, plotted here for a $1.4 M_\odot$ giant star that is $50 R_\odot$ at the onset of mass loss.  The mean density of the star decreases initially upon mass loss then begins to increase again when the bulk of the star's hydrogen envelope is depleted. The response of the star's structure to mass loss determines the impact parameter of the next encounter with the black hole and the quantity of mass removed at pericenter (lower panel). As a result of the changing stellar structure, no single one of the analytic adiabatic response curves can provide an adequate description of the mass loss history. The contours shown are from Equation \eqref{xiad}, shown for $n=1.5 - 2.5$ in intervals of $0.25$ (with $n=1.5$ producing the shallowest profile).   Stars in long orbital periods continue to evolve between passages, which drives the mass transfer episode to completion in fewer orbits. The lines terminate when the star's envelope mass decreases to $5 \times 10^{-3} M_\odot$.   }
\label{fig2}
\end{center}
\end{figure}

Figure \ref{fig2} shows our results for a star that is $1.4 M_\odot$ and $50 R_\odot$ at the onset of mass loss. Due to the giant star's isentropic envelope, it first becomes less dense upon losing mass, resulting in a runaway process in which successive encounters are increasingly disruptive. The lower panel shows the corresponding $\Delta M$ at each pericenter passage. Eventually, when much of the star's hydrogen envelope has been stripped,  the core becomes the dominant gravitational force in the star's structure, and the mean density of the object increases again with subsequent mass loss episodes. This quenches the runaway mass loss, and $\Delta M$ decreases in subsequent passages. Stars in longer orbital periods lose mass with decreasing adiabaticity. These stars evolve farther up the giant branch each orbit, resulting in a slow increase in the hydrogen shell-burning luminosity at the core-envelope interface. This evolution drives these stars to undergo stronger encounters with the SMBH and leads to their envelopes being stripped in fewer orbital periods. 

Also in Figure \ref{fig2}, we make a direct comparison with the mass loss history that would be realized for the adiabatic evolution of a nested polytrope of $1.4 M_\odot$ and $50 R_\odot$ with a $0.3 M_\sun$ condensed core. Here we compute the star's mass-radius relation as $R_*/R_0 = (M_*/M_0)^{\xi_{\rm ad}}$, where $\xi_{\rm ad}$ is given by an approximate formula from \citet{Hjellming:1987ci},
\beq\label{xiad}
\xi_{\rm ad} \approx \frac{1}{3-n} \left(  1-n + \frac{m_{\rm c}}{M_*-m_{\rm c}}\right).
\eeq
The contours for $n=1.5-2.5$ in intervals of $0.25$ reveal that this expression does not provide a reasonable description for the mass loss history from the star. $n=1.5$ corresponds to a convective envelope, higher $n$ indicate an increasing degree of radiative transport in the giant's envelope. The stellar evolution models are therefore essential to capture the realistic adjustment of the mass-losing star. Equation \eqref{xiad} provides a poor fit for two primary reasons. First, the assumption is that the readjustment of the stellar structure is adiabatic. However, through the differences between the {\tt MESA} mass loss histories with different orbital periods, we see that this is not the case. Second, this expression assumes that the giant's envelope has constant $n$ as a function of time, whereas in the stellar evolution models the radial extent of convective and radiative zones evolves as the star loses mass. 

\subsection{Return to the Black Hole}

For each portion of mass removed from the star, about half, $\Delta M / 2$, returns to the black hole. The other half is ejected on hyperbolic orbits. This is formally true if the initial orbit is parabolic, but we will demonstrate that this approximation is reasonable for a wide range of orbital parameters. 
The accretion rate onto the black hole, $\dot M$, is then determined by the rate at which stellar material falls back to the vicinity of the black hole \citep{Rees1988}, 
\beq\label{mdot}
\dot M \approx \dot M_{\rm peak} \left( \frac{t}{\tpeak}\right)^{-5/3}, 
\eeq
where, by requiring that the  $\int \dot M dt = \Delta M /2$,
\beq\label{mdotpeak}
\dot M_{\rm peak} \approx {1\over 3} {\Delta M \over \tpeak}, 
\eeq
and the time of peak, $\tpeak$,  is similar to the fallback time of the most bound material,
\beq\label{tfb}
\tpeak \sim \tfb \approx 120  \left(\frac{\Mbh}{10^7 M_\odot}\right)^{1/2} \left(\frac{M_*}{M_\odot}\right) \left(\frac{R_*}{50 R_\odot}\right)^{3/2}  \text{yr}.
\eeq
This formulation, which treats encounters as nearly parabolic, is a good approximation as long as the spread in energy across the star at pericenter is large compared to the star's orbital binding energy, satisfied for 
\beq\label{arh}
{a \over r_{\rm h} }  \gtrsim 5 \times 10^{-3}  \left(\frac{\Mbh}{10^7 M_\odot}\right)^{0.123} \left(\frac{M_*}{M_\odot}\right)^{2/3} \left(\frac{R_*}{50 R_\odot}\right),
\eeq
where $a$ is the star's orbital semi-major axis and $r_{\rm h}$ is the radius of the black hole sphere of influence, Equation \eqref{rh}. The resultant scalings thus derive partially from the definition of $r_{\rm h}$, which is based on the SMBH $M-\sigma$ relation and discussed in Section \ref{sec4}. 
Another condition for these expressions to be applicable is that the viscous accretion time should be short relative to $\tfb$.
For mass to be stored in a reservoir at its circularization radius, $\sim 2 \rt$, for longer than $\tfb$ (Equation \ref{tfb}), the viscosity would have to be extremely low. 
Taking typical values, the ratio of the viscous accretion timescale to the fallback timescale is 
\beq
{\tau_\nu \over  \tfb} \approx 3 \times 10^{-3} \left({\alpha_\nu \over 10^{-2}}\right)^{-1}  \left({ \Mbh \over 10^7 M_\odot} \right)^{1/2} \left({M_\ast \over 1M_\odot} \right)^{-1/2}  \left(  \frac{H}{R} \right)^{-2}, 
\eeq
where we have assumed an $\alpha$-viscosity disk  \citep{Shakura:1973uy}.

In some cases, the condition expressed in Equation \eqref{arh} is not satisfied. For these more bound orbits, both tidal tails may be bound to the black hole and the most bound material falls back to the SMBH extremely rapidly \citep{Hayasaki:2012ec}. In these cases, the nearly impulsively assembled disk is accreted as mediated by viscosity.  The timescales and temporal evolution of this accretion are complex and remain a subject of debate. For example, see the self-similar solution of \citet{Cannizzo:1990hw} as compared to recent numerical studies by \citet{MontesinosArmijo:2011bl} and \citet{Shen:2013vo}.  The common finding of these studies is that turbulent disk viscosity spreads the accretion of material out over longer timescales that pure fallback. The resulting median accretion rates are thus closer to the average accretion rate than in the pure fallback case. For simplicity, we adopt $\dot M = \langle \dot M \rangle = \Delta M / \torb$ such that a pile-up of material in the accretion disk does not occur between subsequent stellar orbits. However, in Section \ref{sec4} we argue that mass transfer from giant stars in such tight orbits is probably rare due to the destructive effects of direct stellar collisions.

\begin{figure}[tbp]
\begin{center}
\includegraphics[width=\columnwidth]{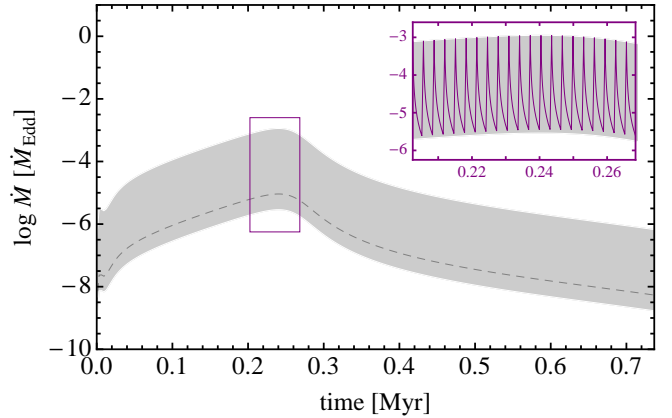}
\caption{ Profile of a repeating flaring episode due to the episodic mass transfer of a giant star to the SMBH. The figure shows a flare from a $1.4M_\odot$, $50R_\odot$ star in a $10^{3.5}$ year orbit about a $10^7M_\odot$ black hole. The gray shaded region shows the overall envelope of the flaring episode, while the dashed line shows the median accretion rate. The inset, with axes in the same units as the main figure, shows that the entire envelope is made of individual flaring episodes with $t^{-5/3}$ decay tails.  }
\label{fig3}
\end{center}
\end{figure}

Figure \ref{fig3} applies the flaring model described by Equations \eqref{mdot}, \eqref{mdotpeak}, and \eqref{tfb} to a $1.4M_\odot$ star orbiting a $10^{7}M_\odot$ black hole with an orbital period of $\torb = 10^{3.5}$ years. The star begins to lose mass to the black hole when it reaches $50 R_\odot$. The shaded region in the main panel shows the overall envelope of the flaring event and the dashed line shows the median $\dot M$. The inset panel reveals each flare to be made up of a short peak and power-law decay phase. Because the star interacts with the black hole only once per orbit at each pericenter passage, the orbital period sets the repetition timescale for the individual flaring episodes. The total duration of the repeating flare is several hundred orbital periods, or approximately $10^6$ yr.

The remnants of spoon-feeding episodes are giant stars stripped of their hydrogen envelopes. The cores of these objects are white dwarfs of helium or carbon/oxygen composition, depending on the mass of the original giant star. Because the white dwarf core is relatively immune to the SMBH's tidal field, this population of remnants are not readily destroyed by the black hole. This population of objects may eventually circularize through dissipation of orbital energy in tides or gravitational waves \citep{Frank:1976tg,Rees1988,Khokhlov:1993bj}. 
This possibility emphasizes the still poorly constrained role that the SMBHs may play in shaping the stellar populations that surround them.

\section{ Estimating the population of mass-transferring stars}\label{sec4}

In this section, we estimate the orbital phase space and number of actively mass-transferring giant stars in a stellar cusp surrounding the SMBH. \citet{Syer:1999gp} have considered the rate at which stars evolve to reach their tidal disruption radius as a component of the tidal disruption rate. This approach is problematic because, as we have shown, stars that evolve to transfer mass to the black hole do so over many orbital periods, rather than a single, fully disruptive encounter.  With the context of repeating flares in mind, we first outline a simple stellar cusp model and the relevant timescales of the stellar dynamical system. We then outline the phase space in which giant stars might be expected to evolve to  transfer mass to the SMBH upon reaching their "loss cone" in angular momentum space \citep{Lightman:1977hu}. 

In what follows, we will use a simplified model of a nuclear star cluster consisting of a power-law stellar density profile $\nu_*(r) \propto r^{-\alpha}$, normalized such that there are a black hole mass of stars within the black hole's sphere of gravitational influence,
 \beq\label{rh}
r_{\rm h} = \frac{G M_{\rm bh}}{\sigma_{\rm h}^2}  = 5.16 \left(\frac{M_{\rm bh}}{10^7 M_\odot }\right)^{0.54} \ {\rm pc},
\eeq
where $\sigma_{\rm h}$ is the external velocity dispersion of the greater galactic bulge. In the numerical expression above we use the $\Mbh-\sigma$ relation \citep[e.g. ][]{Ferrarese2000,Gebhardt2000,Tremaine:2002hd,Gultekin:2009hj,Kormendy:2013vg}, with fitting values from \citet{Kormendy:2013vg}, 
$\sigma_{\rm h} = 2.3 \times 10^5 (M_{\rm bh}/M_\odot)^{1/4.38}  \ {\rm cm \ s^{-1}}$. Interior to $r_{\rm h}$, the black hole is the dominant gravitational influence on stellar orbits, while outside $r_{\rm h}$, stellar orbits are primarily determined by the collective gravitational influence of all of the other stars.

Observations of the centers of early-type galaxies with the {\it Hubble Space Telescope (HST)} have shown that the stellar surface brightness profiles in these galactic centers are bimodal \citep{Faber:1997fn}. Some galaxies exhibit a `cuspy' core that is defined by a power-law rise in surface brightness to the resolution limit of the {\it HST} imaging. Others exhibit a shallower `core' profile with a break radius that is typically similar to the inferred black hole sphere of influence \citep{Faber:1997fn}. Given these observed stellar distributions, one could, in principle, analyze the populations of mass-transferring giant stars expected in those galaxies and compare to their SMBH activity on a case by case basis. As a simple first step, we instead illustrate the expected populations for two representative stellar profiles, $\nu_*(r) \propto r^{-\alpha}$ with $\alpha = 2$ to represent cuspy galactic center profiles and $\alpha=1$ to represent the shallower core profiles \citep[for an illustration of the stellar density profiles, see Figure 12 of][]{MacLeod:2012cd}. These profiles were chosen to capture two extreme cases for the stellar distribution in observed galactic centers in order to illustrate the range of possibilities for the rates of mass-transfer interactions. 

Stars in galactic nuclei live in orbits of period $\torb$ whose energy and angular momentum change on timescales $\te$ and $\tJ$, respectively. $\te$ is the cluster's local relaxation time \citep{Binney2008,Alexander:2005ij}. 
The angular momentum of loss cone orbits, $J \sim \Jlc \approx \sqrt{2 G \Mbh \rt}$, is typically much less than the circular angular momentum for a given orbital semi-major axis $a$, $\Jc \approx \sqrt{ G \Mbh a}$. In other words, $\rp \approx \rt \ll a$, and typical loss cone orbits are very eccentric. 
Thus, the angular momentum change time for orbits that approach the loss cone,
\beq
\tau_J \equiv \left(\Jlc / \Jc\right)^2 \tau_\e,
\eeq
is much less than that for energy, $\te$, because the loss-cone angular momentum is much less than the circular angular momentum, $\Jlc \ll \Jc$.  

Stars are also susceptible to collisions after a time $\tcoll \equiv \torb / N_{\rm coll}(\torb)$, where the number of collisions per orbit is an integral along the orbital path,
\beq
 N_{\rm coll}(\tau_{\rm orb}) = 2 \int_{\rp}^{2a(\e)} \Sigma(s) \nu_*(s) ds ,
\eeq
where $\Sigma$ is the collision cross section of a star, approximately the geometric cross-section $\Sigma \approx \pi R_*^2$ in the high velocity dispersion central regions of the cluster.  For nearly radial loss-cone orbits and power-law density profiles, collisions are dominated by the orbital pericenter approach if $\alpha >1$. If there were to be a break in the power-law profile at small radii, the number of collisions would therefore be reduced. Here we will also assume that collisions are always destructive, even though this may not be the case for high velocity collisions involving giant stars \citep{Bailey:1999bs,Dale:2009ih}. In this way, we calculate an upper limit to the collisional destruction of giant stars that might otherwise go on to spoon-feed the black hole. 

Finally, the tidal radius of a star changes as a function of time due to stellar evolution according to a timescale 
\beq
\tau_{\rm evol} \equiv \frac{\rt}{{\dot r}_{\rm t}} \approx \frac{R_*}{\dot R_*}.
\eeq
$\tevol$ is long on the main sequence and shortens as the star evolves off the giant branch. Because this change happens over the course of the red giant branch, this timescale is roughly the length of the red giant branch, $\langle \tevol \rangle \sim \tau_{\rm rgb}$. We adopt $\tevol = 10^7 \text{ yr}$ in the following analysis.

\begin{figure}[tbp]
\begin{center}
\includegraphics[width=\columnwidth]{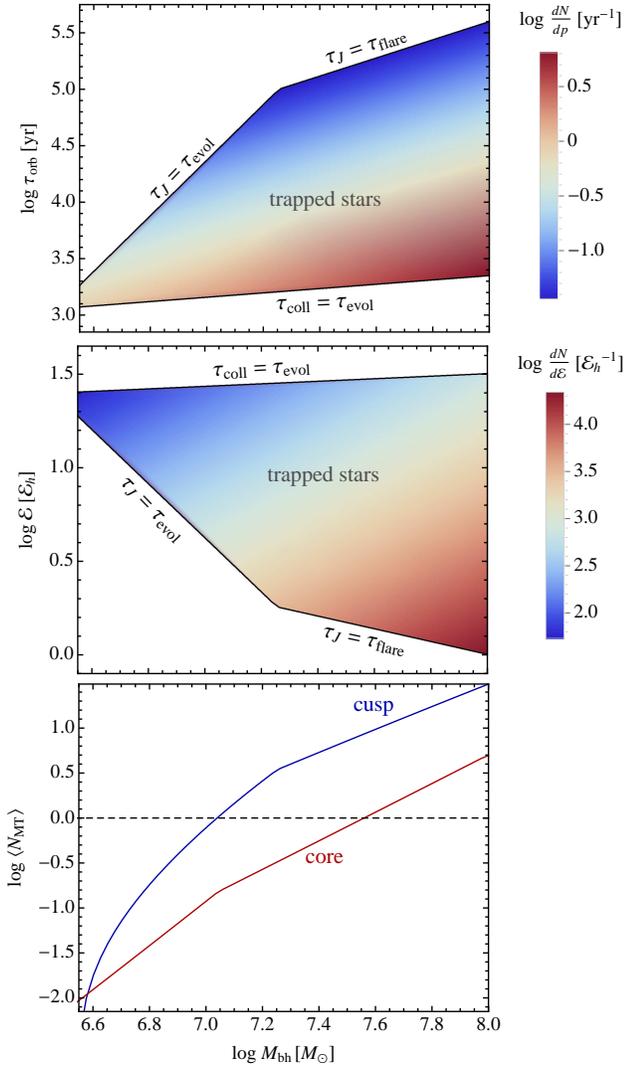}
\caption{Phase space of stars that can evolve to transfer mass to the SMBH. The top two plots assume a cuspy stellar density profile  $\nu_*(r) \propto r^{-2}$, as does the blue line in the third panel, while the red, core, line in the lower panel takes $\nu_*(r) \propto r^{-1}$ \citep[e.g.][]{Faber:1997fn}. The phase space for tightly bound stars is limited by collisions. Two considerations limit the more weakly bound population: first, that the star reach the loss cone through evolution rather than under the influence of a random walk in angular momentum, $\tevol < \tJ$, and, second, that once mass transfer begins, the flaring event transpires more rapidly than the random walk timescale, $t_{\rm flare} <\tJ$.  The lower panel shows the number of mass-transferring stars that might be expected at any given time,  estimated as $\langle N_{\rm MT} \rangle= 10^2 \langle \tau_{\rm orb} \rangle \Gamma_{\rm evol}$, where  $\langle \tau_{\rm orb} \rangle$ is the orbital period averaged over the distribution of stars in binding energy.  }
\label{fig4}
\end{center}
\end{figure}

The population of stars that can evolve to reach the loss cone is bounded to have angular momenta that are low enough that the black hole's tides will impinge on the star at some point during the star's giant-branch evolution, thus
\beq\label{Jcondition}
J < \Jlc(M_*, R_{\rm max} ) ,
\eeq
where $R_{\rm max}$ is the maximum stellar radius reached during the red giant phase. This condition implies $\rp < \rt(M_*, R_{\rm max} )$. 
This population of stars is also is bounded in orbital energy space by several considerations based on the timescales of various drivers of evolution in the star's orbital parameters. The first condition is that stellar evolution, rather than the star's orbital random walk, must drive the star to reach the loss cone, $\tevol < \tJ$. Second, the orbital random walk must be small during the flare duration, $\tau_{\rm flare} < \tJ$. Because the peak of the flaring event takes $\sim 10^2$ orbits, we make the approximation $\tau_{\rm flare}\approx 10^2  \torb$. These conditions define the least bound objects, which have shorter $\tJ$. The most bound objects exhibit very nearly Keplerian orbits, but since they pass through the densest central regions of the stellar cusp $\tevol/\torb$ times during their giant branch lifetime, they may be extremely vulnerable to collisions. Figure \ref{fig4} shows these constraints on the phase space of trapped stars in the upper two panels. 

The shading in Figure \ref{fig4} shows the distribution of orbits in orbital energy $\e$, and in orbital period $\torb$.  
These distributions are computed by considering the number of stars per unit energy that satisfy the low angular momentum condition of Equation \eqref{Jcondition}. This number is $dN/d\e = 4 \pi^2 f(\e) \torb(\e) \Jlc(M_*, R_{\rm max} )$, where $f(\e)$ is the distribution function of stars in orbital energy \citep{Magorrian:1999fd}. Following \citet{Magorrian:1999fd} we have assumed that the distribution function is isotropic in $J^2$ and therefore only depends on energy. The distribution function has scaling $f(\e) \propto \e^{\alpha - 3/2}$, while from Kepler's law, $\torb(\e) \propto \e^{-3/2}$. As a result, $d N / d \e \propto \e^{\alpha -3}$ \citep{Magorrian:1999fd}. Thus, for an $\alpha = 2$ stellar density profile, orbits within the loss cone are distributed in energy as $d N / d \e \propto \e^{-1}$. The total number of stars trapped in this phase space that will lead to mass transfer with the SMBH is then an integral over $d N / d \e$, 
\beq
N_{\rm trapped} = \int_{\e_{\rm min}}^{\e_{\rm max}} \left( \frac{dN}{d\e} \right) d \e, 
\eeq
where $\e_{\rm min}$ and $\e_{\rm max}$ are limits on the energy based on the comparisons of timescales described above. 
For $\alpha =2$, different orbital energies contribute equally to the integrated number of stars trapped within the giant star loss cone, because $d N / d \e \propto \e^{-1}$ or $d N / d \log \e = \text{constant}$. 
These flares have equal likelihood of occurring with any pericenter distance because of the isotropic angular momentum distribution of the stellar cluster,  thus $dN/dJ^2$ and  $dN/d\rp$ are both constant. These flat distributions imply that a mass transfer episode involving a star of a given mass between 10 and 11 $R_\odot$ is equally likely as when that star is between 100 and 101 $R_\odot$. The average stellar radius on encountering the SMBH is therefore $\approx R_{\rm max}/2$, where $R_{\rm max}$ is the maximum radius of stars on the red giant branch. 

The rate at which the $N_{\rm trapped}$ trapped stars evolve to reach the loss cone is given by the mean lifetime of the stars. We find,
\beq\label{evolrate}
\Gamma_{\rm evol} \approx \frac{N_{\rm trapped}}{\tau_{\rm life}}  = 10^{-6} \left(\frac{\tau_{\rm life}}{1 \ {\rm Gyr}} \right)^{-1} \left(\frac{ N_{\rm trapped}  }{10^3} \right) \text{yr}^{-1},
\eeq
where $\tau_{\rm life}$ is the age of the stellar population. By comparing the duration of a typical flare, approximated as $ 10^2  \torb$, with the rate above, we can estimate the number of stars expected to be actively mass transferring with the black hole at any given time. To do so, we must incorporate knowledge of the orbital period distribution of mass-transferring stars. Figure \ref{fig4} shows the distributions of trapped stars in orbital binding energy and period. In the lower panel of  Figure \ref{fig4}, we show the resultant number of actively mass-transferring stars, computed as $N_{\rm MT} = 10^2 \langle \tau_{\rm orb} \rangle \Gamma_{\rm evol}$, based on the average orbital period, $\langle \tau_{\rm orb} \rangle$. Black holes with masses greater than approximately $10^7 M_\odot$ might then be expected to host an actively mass-transferring  population of trapped giant stars. The nuclei of lower-mass black holes are characterized by higher stellar number densities, and thus have shorter relaxation and collision times. In these nuclei $(\Mbh \lesssim 10^7)$ mass transfer events do occur but with a low duty-cycle.

\section{ Significance to the duty cycle of tidally-fed SMBHs}\label{sec5}

In this section, we discuss the role of direct tidal feeding episodes in the duty cycle of low-level SMBH activity.  We illustrate these processes with Monte Carlo realizations of black hole accretion histories. We make use of the stellar cluster properties described in Section \ref{sec4} with $\nu_* \propto r^{-2}$ and assume a stellar mass of $1.4 M_\odot$,   corresponding to a main-sequence turnoff age of approximately $4$ Gyr. While this choice is meant to be illustrative rather than exact, similar stellar ages are seen in the nuclear region of M32 by \citet{Seth:2010kh}. Such stars have a giant-branch lifetime of  $\sim4\times10^8$ years, and thus spend $\sim10$\% of their lifetime on the giant branch. The effect of stellar population age on the rate at which stars evolve to reach the loss cone and begin to transfer mass can be seen in equation \eqref{evolrate}, thus the expected event rate would be lower by a factor of three if the main-sequence turnoff age were 12 Gyr, and higher by a factor of four if it were instead 1 Gyr. 

\begin{figure*}[tbp]
\begin{center}
\includegraphics[width=0.96\textwidth]{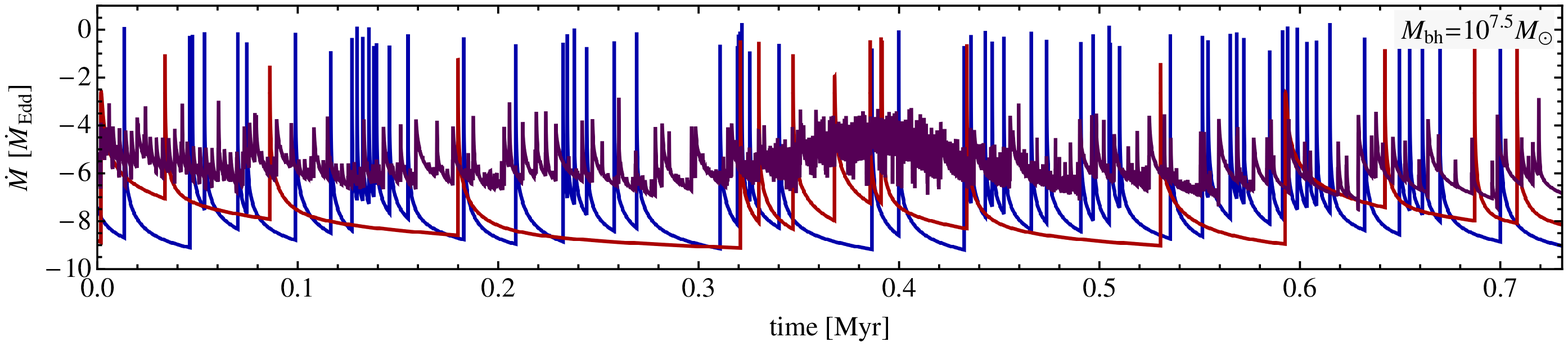}

\includegraphics[height=0.3\textwidth]{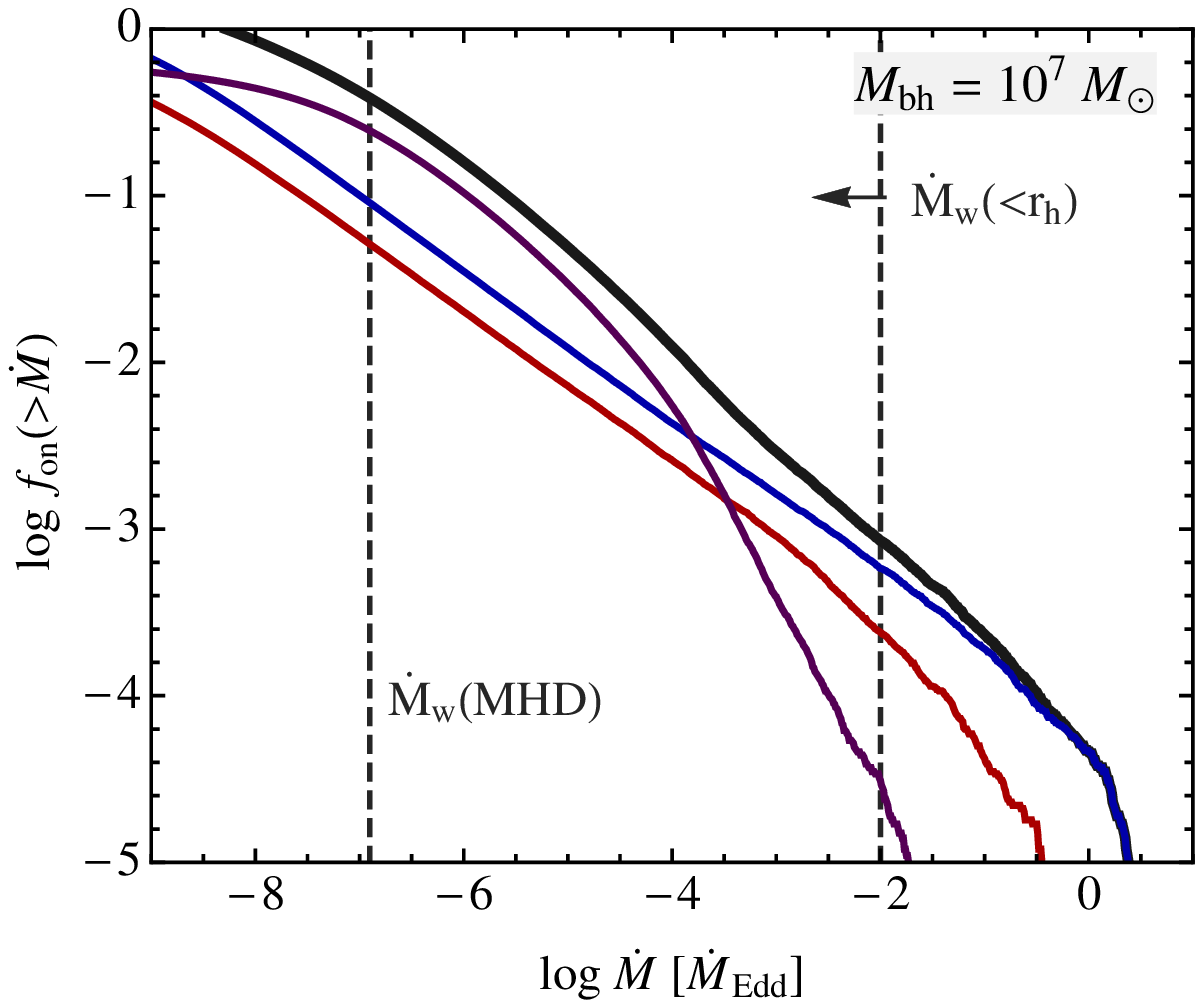}
\includegraphics[height=0.3\textwidth]{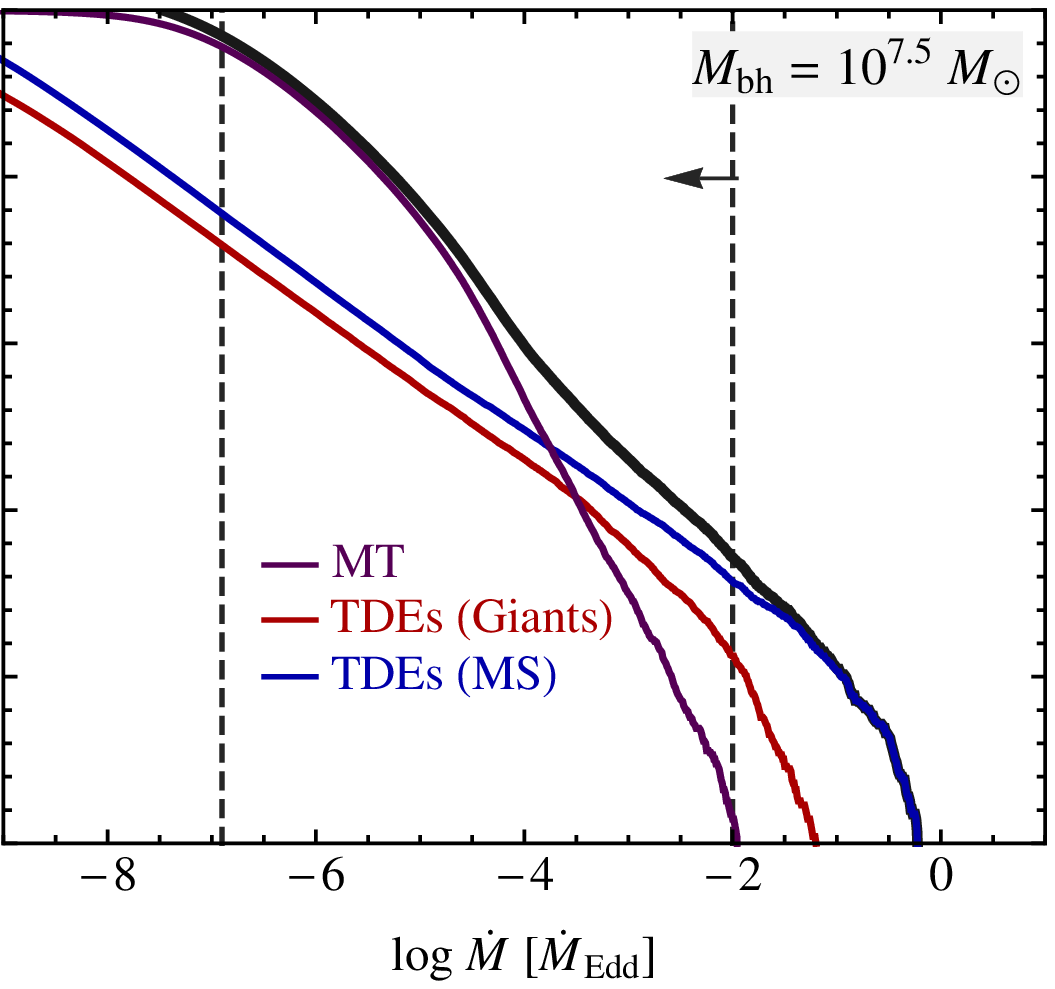}
\includegraphics[height=0.3\textwidth]{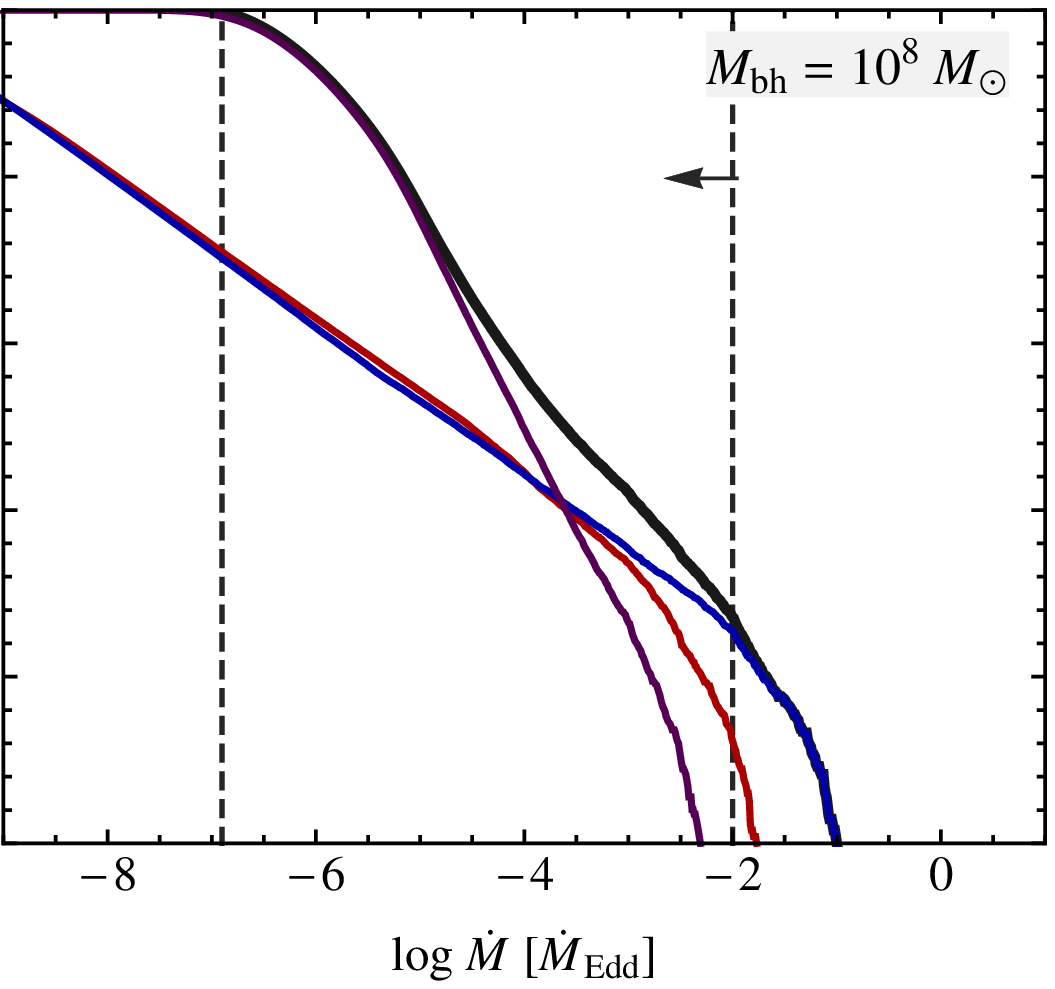}

\caption{Monte Carlo realizations of the duty cycle of tidally fed black hole activity for SMBH masses of $10^{7}M_\odot$, $10^{7.5}M_\odot$, and $10^{8}M_\odot$. The blue line shows the tidal disruption of main-sequence stars, labeled TDEs (MS); the red line shows giant branch stars, labeled TDEs (Giants). These tidal disruption components only include the fraction of events fed from the large angle scattering regime where $\Delta J > \Jlc$. A fraction $\rs/\rt$ of events are promptly swallowed by the SMBH, here we plot only those that pass outside $\rs$ and produce a flare. The purple line shows the contribution from episodically mass-transferring giant stars, labeled MT. For black hole masses $\gtrsim 10^7 M_\sun$ the number of actively mass-transferring stars is $\gtrsim1$. These mass transfer episodes dominate above the decay tails of tidal disruption events at low Eddington ratios. We expect these mass-transferring stars to be the dominant contribution to tidally fed SMBH activity at $\dot M/\dot M_{\rm Edd} \lesssim 10^{-4}$. The dashed lines show two estimates of the degree to which stellar winds may feed the SMBH. The total stellar wind injection with the sphere of influence is an upper limit and is marked as $\dot M_w (<r_{\rm h})$. The winds found to actually accrete in models of Sgr A* are a small fraction of that, $\dot M_w (MHD)$ from simulations of the accretion flow by \citet{Shcherbakov:2010gs,Shcherbakov:2013ip}. 
$\dot M/ \dot M_{\rm Edd}$ is computed assuming $\dot M_{\rm Edd} = 0.02 (\Mbh/10^6 M_\odot) M_\odot \text{yr}^{-1}$, which corresponds to a radiative efficiency of $\eta = 0.1$.}

\label{fig5}
\end{center}
\end{figure*}

Accretion histories (a subset) and duty cycles are shown for $\Mbh = 10^7 M_\odot$, $10^{7.5} M_\odot$, and $10^8 M_\odot$ in Figure \ref{fig5}. 
We draw the stellar parameters of tidal disruption flares according to the methods described in \citet{MacLeod:2012cd} and scale the profiles derived from hydrodynamic simulations in timescale and $\dot M$. The relative likelihood of disruption of different stellar types scales with their occurrence in the stellar population and $\rt^{1/4}$ \citep{Wang:2004jy,Milosavljevic:2006jj,MacLeod:2012cd}.
We draw the giant star mass transfer events from the same stellar cluster distribution. 
 Using the distributions in Figure \ref{fig4}, we populate the stars' orbital periods and assume that the mass transfer profile follows one of the profiles from Figure \ref{fig2}, choosing the nearest in logarithmic space to the orbital period. We then compare the resultant feeding directly to observed Eddington ratio distributions in local quiescent galaxies and to the indirect SMBH feeding by stellar winds.

In this section, we normalize accretion rates with respect to black hole mass to a fiducial Eddington accretion rate $\dot M_{\rm Edd} = 0.02 (\Mbh/10^6 M_\odot) M_\odot \text{yr}^{-1}$, which corresponds to a radiative efficiency of $\eta = 0.1$ where $L = \eta \dot M c^2$ and thus $\dot M_{\rm Edd} =   L_{\rm Edd} \eta^{-1} c^{-2}$. The normalized mass accretion rates thus show what the bolometric accretion luminosity, $L/L_{\rm Edd}$, of a SMBH would be given a certain radiative efficiency, and may be scaled to different values of the radiative efficiency according to $(\eta/0.1)$.

\subsection{Tidal disruption flares}\label{sec51}

 At the highest Eddington ratios, main-sequence flares dominate the SMBH feeding (Figure \ref{fig5}). At lower accretion rates main-sequence and giant-star flares contribute similarly to the duty cycle despite the lower rate of giant-star tidal disruption events because the decay timescale $\tfb$ is longer for giant-star disruptions \citep[Equation \eqref{tfb}, and][]{MacLeod:2012cd}. 
The late-time power law decay tails of tidal disruption events $(\sim t^{-5/3})$ give rise to the power law seen in the duty cycle at lower Eddington ratios. In particular if the late time decay follows Equation \eqref{mdot}, then the duty cycle can be written 
\beq\label{dutycycle}
f_{\rm on}(> \dot M) = \frac{\tfb }{t_{\rm repeat}} \left(  \frac{\dot M}{\dot M_{\rm peak}} \right)^{-3/5}, 	
\eeq
where $t_{\rm repeat}$ is generally the inverse of $\Gamma_{\rm TDE}$, the rate of tidal disruption events. While the simulation results used in Figure \ref{fig5} differ slightly in late-time power law slope, these basic scalings give intuition for the low Eddington ratio duty cycle that results from tidal disruption events. One consequence is that since typically $\tfb \ll t_{\rm repeat}$, the accretion rate at which $f_{\rm on} \sim 1$ is far less than $\dot M_{\rm peak}$. In Figure \ref{fig5}, the steady state accretion rate between flares (for which the duty cycle is of order unity) is only achieved at $\dot M \sim 10^{-9} \dot M_{\rm edd}$. A very similar analysis has been performed by \citet{Milosavljevic:2006jj} who further show that the resulting luminosity function may explain $\sim 10$\% of local AGN activity. While our analysis is in agreement with previous work that has shown that tidal disruption flares cannot explain the entire local AGN luminosity function, we do find that quasi-steady feeding from mass-transferring giant stars may provide a significant contribution to the median accretion rate between luminous flaring episodes. 

\subsection{Episodic mass transfer from evolving stars: Spoon-feeding}
Giant stars that grow through stellar evolution to  episodically transfer mass at pericenter over the course of many orbital periods to the SMBH. As a result, the SMBH feeding from this spoon-fed material is smeared over long timescales by the episodic nature of the individual flaring episodes.  Mass-transferring stars therefore generally feed the SMBH at low Eddington ratios but at quasi-steady rates. This can be seen mathematically through inspection of Equation \eqref{dutycycle}. Mass transfer episodes repeat every orbital period, reducing $t_{\rm repeat}$, and therefore achieve a duty cycle of order unity at higher $\dot M$ than full tidal disruption flares. These events can be expected to establish a minimum accretion rate above that established by the tails of tidal disruption events in systems where the number of actively mass transferring stars (the lower panel of Figure \ref{fig4}) is of order unity or greater, $\langle N_{\rm MT} \rangle \gtrsim 1$. In systems where $\langle N_{\rm MT}  \rangle < 1$ stars do occasionally grow to the loss cone and feed the black hole, but they are not able to establish a quasi-steady floor accretion rate with $f_{\rm on} \sim 1$.  

Episodic flares from evolving stars are represented with the purple line in Figure \ref{fig5}. In the upper panel, we show a representative timeseries of accretion rate to the SMBH, several interesting features appear. First, it is obvious that while they never reach luminosities similar to the peaks of tidal disruption flares, the mass transfer from spoon-fed giant stars does fill in the "gaps" between tidal disruption flaring events. The quasi-steady accretion rate that results ranges between $\dot M \sim 10^{-4} - 10^{-6} \dot M_{\rm Edd}$. The structure in the curve results from the modulation of repeating flares of different orbital periods. Stars in relatively long period orbits have longer time between episodes and relatively deeper troughs between peaks. This overall curve is mediated by shorter timescale variation from flares with shorter repetition times. At some times, like near $0.4$ Myr, it is possible to see the characteristic structure of an entire flaring episode with relatively short orbital period. Here the flare and decay cycles are so rapid that they appear blurred in the Figure, and instead we see the shape of the overall flaring envelope, as in Figure \ref{fig3}. 

The dominance of mass transfer episodes over the tails of tidal disruption events can be seen in the lower panels of Figure \ref{fig5}. The relatively steep profile of these duty-cycle curves arises from the small range of $\dot M$ achieved in spoon-feeding events, as seen both in the upper panel of Figure \ref{fig5} and the single event shown in Figure \ref{fig3}. 
We find that mass-transferring stars are the dominant contribution to tidally fed SMBH activity at $\dot M\lesssim 10^{-4} \dot M_{\rm Edd} $, particularly for relatively massive black holes $\Mbh \gtrsim 10^7 M_\odot$. A similar behavior is realized at slightly higher black hole masses in the case of galactic nuclei with shallow cores. In these galactic centers, spoon-fed stars might be expected to make a meaningful contribution to the duty cycle in systems with $\Mbh \gtrsim 10^{7.5} M_\odot$. This behavior can  be inferred directly from the lower panel of Figure \ref{fig4}, in which we show the averaged duty cycle for the two extreme cases of the galactic nucleus structure.   As discussed in Section \ref{sec4}, shallower core-like profiles are probably most relevant for massive black holes $\Mbh \gtrsim 10^8 M_\sun$, thus we expect that spoon-fed giant stars will feed these very massive black holes but according to the core galaxy duty cycle shown in Figure \ref{fig4}. SMBHs with mass $\Mbh \gtrsim 10^8 M_\sun$ will swallow main sequence stars whole rather than tidally disrupting them \citep[e.g.][]{MacLeod:2012cd}, but giant stars might still be expected to feed the SMBH through the spoon-feeding and tidal disruption channels.

\subsection{Diffusion to the loss cone}
The gradual diffusion of main-sequence stars in angular momentum space to the loss cone will have little effect on the SMBH activity duty cycle.  In general, these events contribute a fraction of the flux of stars into the loss cone. They will not result in a single, strongly disruptive passage by the SMBH. 
Only objects that are fully convective such as the lowest mass $(M\lesssim 0.4 M_\odot)$ and very high mass $(M\gtrsim 20 M_\odot)$  main-sequence stars exhibit a mass-radius relationship that allows for a runaway response to mass loss over multiple orbits. Sun-like objects (which may be modeled by $n=3$ polytropes) exhibit a strongly protective initial response to mass loss \citep{Hjellming:1987ci}. Further, the critical radius that defines the transition between scattered orbits and diffusing orbits moves to tighter binding energies for stars with small $\Jlc$ like main-sequence stars \citep{Frank:1976tg,Lightman:1977hu}. The orbital periods of these objects become short (thus they contribute less to the duty cycle of SMBH activity). Additionally, in cuspy nuclei that are dynamically relaxed, mass segregation likely leads to a relatively small number of low mass stars that are very tightly bound \citep[e.g.][]{Alexander:2005ij}. Secular effects like the Schwarzschild barrier also become important for such tightly bound orbits \citep{Merritt:2011ea}. Main-sequence stars that are partially disrupted by the SMBH are likely kicked out by the resultant mass loss asymmetry, which increases their specific orbital energy by a factor $\sim G M_*/R_*$ which is  greater than the orbital binding energy for typical orbits when evaluated for the main-sequence mass and radius \citep{Manukian:2013ce}.

Giant branch stars, if they are able to diffuse to reach the loss cone, likely undergo a spoon-feeding encounter history similar to that described in this paper. The population of objects able to do so may be limited. Collisions are particularly damaging for giants in galactic nuclei \citep{Davies:1998je,Bailey:1999bs,Dale:2009ih} and may be even more severe for giants on very eccentric orbits that pass through the densest regions of the stellar cluster every orbit \citep{MacLeod:2012cd}.
In high mass systems, $\Mbh \gtrsim 10^7 M_\odot$, evolution to the loss cone dominates, as shown in Figure \ref{fig4}. 
In systems with lower mass black holes, $\Mbh  \sim 10^6 - 10^7 M_\odot$, diffusion to the loss cone may be possible. 
Only a few percent of diffusing stars that reach the loss cone will be giants because in the diffusion limit there is only a logarithmic enhancement in the loss cone rate with tidal radius, $\Gamma_{\rm diff} \propto \text{ln}(\rt)$ \citep{Lightman:1977hu}. A careful consideration of the importance of these stars would need to include the orbital random walk in the stars' mass transfer histories and is an interesting future application of the episodic mass transfer model presented in this paper. 

\subsection{Comparison to stellar wind feeding}
Stellar winds feed material indirectly to SMBHs by ejecting material into the cluster medium \citep[e.g.][]{Holzer:1970eo}. Simulations of stellar wind feeding onto massive black holes typically find two key features of the accretion flow. First, only a small amount of the material ejected into the cluster medium ever reaches the SMBH. The low inflow rate arises because winds inject both mass and kinetic energy into the cluster.  Young star winds, in particular, may have super-virial velocities, leading to the ejection of the bulk of the mass \citep[e.g.][]{Quataert:2004bp,Rockefeller:2004da,Cuadra:2005hu,Cuadra:2006ju,Cuadra:2008bd,DeColle:2012bq}. Second, the accretion flows tend to have low net angular momentum because they result from the combined contribution of many stars. In a spherical nucleus almost perfect cancelation of angular momentum results, and the accretion flow circularizes only at very small radii \citep{Cuadra:2008bd}. The low angular momentum of the accretion flow may contribute to the inferred low radiative efficiency of accretion fed by stellar winds \citep{Baganoff:2003ck,Ho:2009fq}, particularly if the circularization radius is inside the transition to an advection dominated accretion flow (ADAF) \citep{Fabian:1995ur,Narayan:1998ef,Quataert:1999kc,Blandford:1999dp}. 

In Figure \ref{fig5}, we include several direct comparisons to models of stellar wind feeding. \citet{Shcherbakov:2010gs} and \citet{Shcherbakov:2013ip} use magnetohydrodynamic models of the accretion flow onto Sgr A* to conclude that an inflow rate of $\sim 10^{-8} M_\odot \text{ yr}^{-1}$ is consistent with the feeding. This corresponds to an Eddington ratio $\dot M_{\rm w}(\text{MHD}) / \dot M_{\rm Edd} = 1.25 \times 10^{-7}$ and is shown in the panels of Figure \ref{fig5} as scaling with mass to maintain a constant Eddington ratio. Considerably more mass is available to the SMBH within the nuclear cluster environment. As a strict upper limit, we take a cluster of $1.4 M_\odot$ stars that lose mass at rate $\langle \dot M_* \rangle \approx 0.8 M_\odot / 4 \text{Gyr} = 2 \times 10^{-10} M_\sun \text{ yr}^{-1}$ based on the initial-final mass relation and lifetime of these stars \citep{Kalirai:2008js}. Multiplying by the number of stars within the SMBH's sphere of influence gives a crude upper limit to the mass contributed by stellar winds that is potentially available to the black hole of $\dot M_w(<r_{\rm h}) \lesssim 10^{-2} \dot M_{\rm Edd}$. Considerable uncertainty exists between these two limiting cases, in particular with respect to the degree to which material reaches the black hole, and the radiative efficiency with which it will shine. 
 By contrast, material spoon-fed to the SMBH is injected at small radii (comparable to the tidal radius), making it less susceptible to feedback or outflows as it accretes than are stellar winds, which are dominantly injected at the sphere of influence. 
We see from Figure \ref{fig5} that spoon-feeding from mass-transferring giant stars can lead to  accretion rates within the range of those those implied by stellar wind feeding in galactic nuclei of $\Mbh \sim 10^7 - 10^8 M_\sun$.

\subsection{Implications for low-luminosity active galactic nuclei}\label{sec55}

We compare our predictions about the duty cycle of spoon-fed SMBHs to distributions of $L/L_{\rm Edd}$ from \citet{Ho:2009fq}.
Using data from the Palomar sample of local active galactic nuclei (AGN), \citet{Ho:2009fq} finds that most galactic nuclei shine at a very small fraction of their Eddington limit. 
The range of median luminosities for AGN classes computed by \citet{Ho:2009fq} span several orders of magnitude. From low to high $L/L_{\rm Edd}$ these are Absorption ($2.2\times10^{-7}$), Transition ($1.5\times10^{-6}$), Liner ($6 \times10^{-6}$), and Seyfert ($1.1\times10^{-4}$) nuclei. Some uncertainty lies in the determining whether accretion luminosity is the true source of nuclear activity at such low $L$. Potential sources of contamination include low-mass x-ray binaries \citep{Miller:2012bh} or perhaps even diffuse emmission \citep[e.g.][]{Soria:2006ba}. Thus, the constant H$\alpha$ and x-ray to bolometric corrections assumed by \citet{Ho:2009fq} are suggested to carry an error for individual sources $\lesssim 0.7$ dex.  
\citet{Ho:2009fq} uses these data to conclude that we must be primarily observing the signatures of radiatively inefficient $(\eta \ll 1)$ accretion of a relatively large amount of material supplied by stellar winds. However, there remain large uncertainties in the degree to which gas within the black hole sphere of influence is able to accrete (due to either feedback processes or outflows). Thus, both the radiative efficiency, and the efficiency with which material reaches the SMBH are parameterized in this conversion from $\dot M$ to $L$. 

At low black hole masses, $\Mbh \lesssim 10^7 M_\sun$, stellar winds likely set the minimum accretion floor in gas-deprived galaxies. For higher black hole masses in the range $\Mbh \sim 10^7 - 10^8 M_\sun $ plotted in Figure \ref{fig5}, we suggest that the inferred $L/L_{\rm Edd}$ Eddington ratios in some local galactic nuclei may also be consistent with the digestion of mass from spoon-fed stars given efficiencies $\eta \sim 10^{-3}-10^{-1}$. Because stellar wind feeding and direct tidal feeding of SMBHs potentially result in different morphologies of the resulting accretion structure, one might not necessarily expect that $\eta$ is fixed for a given fuel supply, $\dot M$. A more complete picture of the properties of the lowest luminosity AGN may provide some information about their fueling mechanism, especially when coupled with more detailed modeling of the accretion flows themselves. In particular, parallel constraints on the activity level of SMBHs and the stellar distributions that surround them may offer insight into the feeding mechanism and, in turn, the morphology and radiative efficiency of the resulting accretion flow.

\section{Conclusions and Future Work}\label{sec6}

To understand of the nature of the accretion flows in quiescent galactic nuclei, we must make certain our census of potential fuel sources is complete. To this end, we study the SMBH feeding that arises when stars trapped in eccentric orbits evolve to  transfer mass episodically to the SMBH. We call this process spoon-feeding giant stars to SMBHs. About half the mass stripped from the star falls back towards the SMBH, lighting it up in an accretion flare. The remnant returns to the SMBH to transfer mass once per orbital period. We show that the thermal evolution of the remnants determines the magnitude of the subsequent mass-loss episodes. We compute orbital histories in which we self-consistently compute the adjustments to the the stellar structure in the {\tt MESA} stellar evolution code. A typical low-mass giant branch star may transfer mass for $\sim 10^2$ orbits before leaving behind a helium white dwarf remnant. We estimate that a steady-state population of these mass-transferring stars is likely to exist in galactic nuclei hosting SMBHs more massive than approximately $10^7 M_\odot$. In nuclei with lower density cores, this transition happens at somewhat higher black hole mass $\Mbh \gtrsim 10^{7.5} M_\odot$. Using Monte Carlo realizations of SMBH accretion histories, we show that this population of mass-transferring stars contributes significantly to the duty cycle of low-level SMBH activity. The feeding from these stars may exceed that from the decay tails of tidal disruption events at $\dot M \lesssim 10^{-4} \dot M_{\rm Edd}$.

In this work, we have presented a preliminary formalism for modeling the mass transfer that occurs when giant stars in eccentric orbits grow  to reach their loss cone and spoon-feed material to the SMBH. This approach may be extended to explore several interesting questions. In particular, we have considered the idealized case in which the star's orbit does not evolve in the course of the encounter. Stars that diffuse in angular momentum undergo a random walk in angular momentum due to two-body relaxation \citep{Lightman:1977hu}, while stars in non-isotropic structures like rings, disks, or triaxial clusters are subject torques that can drive coherent orbital evolution  \citep{Magorrian:1999fd,Merritt:2010he,Madigan:2009gd,Madigan2011,Vasiliev:2013ua,2013ApJ...763L..10A}. These differing cases of orbit evolution likely imprint themselves on the light curves of the resulting mass-transfer flares in unique ways,  and in turn the duty cycle of SMBH activity,  perhaps offering an avenue to explore the dynamical processes at play in distant galactic nuclei \citep[e.g.][]{Rauch:1998jr}. In the giant star case we have analyzed, the encounter itself does not change the orbital parameters significantly. This is partially because of the star's stratified core-envelope structure \citep{Liu:2012er} and partially because the star's specific orbital energy $\sim G \Mbh/a$ is larger than the star's specific binding energy, $G M_*/R_*$ \citep{Manukian:2013ce}. For main-sequence stars and compact stellar remnants with higher escape velocities, this is not necessarily the case. The orbital energy change imprinted by mass loss becomes an important factor in determining the star's encounter history with the SMBH. 

We further assume in this work that the primary change in stellar structure is due to the change in envelope mass. As we demonstrate in Section \ref{sec2}, this is almost certainly the case for giant stars where surface heating is counteracted by rapid radiative cooling \citep{1987ApJ...318..261M}. However, in other classes of objects, particularly those for which their thermal adjustment time is long compared to an orbital period, the heat deposited into the star may play a dramatic role in modifying the star's structure and mass transfer history with the SMBH. Thus, the additional heat deposited deep within a star may not be able to be radiated effectively in an orbital period. If this is the case, the star will adiabatically expand \citep{Podsiadlowski:1996wv} and more strongly feel the tidal force of the SMBH in its subsequent passages. This process can easily lead to a runaway, as has been found by \citet{Guillochon2011} in the case of giant planets interacting with their host stars. The rotation excited by the encounter with the SMBH is an additional effect on the stellar structure we do not address here. The expected rotation may be a significant fraction of the star's 
breakup rotation, leading to mixing and and other effects that can modify stars' subsequent evolution \citep{Alexander:2001fw}, while also reducing their mean density. This may make stars more prone to disruption in subsequent passages. 

In Section \ref{sec5}, we illustrate how single-mass stellar clusters might imprint their presence on the accretion history and thus also the luminosity function of low-luminosity AGN. 
The stellar cusps around SMBHs are certainly not single stellar mass, but our knowledge of  stellar populations in galactic nuclei other than our own \citep[e.g.][]{Schodel2007} is weakly constrained. The detailed nature of these stellar populations plays a significant role in determining the relative rates of different classes of tidal interactions between stars and SMBHs \citep{MacLeod:2012cd}. As we outline above, further complexities in the orbital distribution and orbital evolution of the population of stars residing close to SMBHs in galactic centers will also imprint themselves on the luminosity function of low-level AGN. Thus, direct comparisons between observed distributions of SMBH activity and predictions based on various feeding mechanisms will offer a statistical window into the more general nature of the stellar dynamics and populations of galactic center stellar clusters.

\acknowledgments{
We are grateful to the anonymous referee and to Martin Rees for comments that improved this work. 
It is a pleasure to thank Fabio Antonini, Elena Gallo, Jacqueline Goldstein, Doug Lin, Brian Metzger, Cole Miller, Jill Naiman, Johan Samsing, Anil Seth, Nick Stone, and Simon Portegies Zwart, as well as the participants of the Snowpac 2013 ``Black Hole Fingerprints'' conference for helpful discussions. 
M.M. benefited from advice and discussions at the 2012 {\tt MESA} summer school and the hospitality of the DARK cosmology centre. 
The software used in this work was in part developed by the DOE-supported ASCI/Alliance Center for Astrophysical Thermonuclear Flashes at the University of Chicago. Simulation visualizations were created with the $\texttt{yt}$ toolkit \citep{Turk:2010dd}. We acknowledge support from the David and Lucile Packard Foundation, NSF grant: AST-0847563, the NSF Graduate Research Fellowship (M.M.), and the NESSF graduate fellowship (J.F.G.). }

\bibliographystyle{apj}

\end{document}